\documentclass[preprint, superscriptaddress]{revtex4-1}

\usepackage{amsfonts}           
\usepackage{amssymb}            
\usepackage{amsmath}            
\usepackage{comment}
\usepackage{latexsym}           
\usepackage{dcolumn}            
\usepackage[dvips]{graphicx}   
\usepackage{enumerate}
\usepackage{epstopdf}           
\usepackage{fancyhdr}           
\usepackage{hyperref}           
\usepackage{setspace}           
\usepackage{xspace}             
\usepackage{subfigure}          
\usepackage{bm}                 
\usepackage[ulem=normalem]{changes}
\usepackage{url}
\usepackage{color}
\newcommand{\bea}{\begin{eqnarray}}
\newcommand{\eea}{\end{eqnarray}}

\newcommand{\cf}{{\it cf.}~}

\newcommand{\etal}{{\it et al.}}

\newcommand{\bfx}{{\bf x}}

\newcommand{\eps}{\epsilon}

\newcommand{\eg}{{\it e.g. }}

\newcommand{\been}{\begin{enumerate}}
\newcommand{\een}{\end{enumerate}}
\newcommand{\beit}{\begin{itemize}}
\newcommand{\eit}{\end{itemize}}

\newcolumntype{L}[1]{>{\raggedright\let\newline\\\arraybackslash\hspace{0pt}}m{#1}}
\newcolumntype{C}[1]{>{\centering\let\newline\\\arraybackslash\hspace{0pt}}m{#1}}
\newcolumntype{R}[1]{>{\raggedleft\let\newline\\\arraybackslash\hspace{0pt}}m{#1}}

\begin{document}


\title{Learning local, quenched disorder in plasticity and other crackling noise phenomena}
\date{\today}

\author{Stefanos Papanikolaou}
\affiliation{Department of Mechanical Engineering, The West Virginia University}
\affiliation{Department of Physics, The West Virginia University}
\affiliation{Department of Mechanical Engineering, The Johns Hopkins University,  Baltimore, MD 21218}

\begin{abstract}
{
When far from equilibrium, many-body systems display behavior that strongly depends on the initial conditions. A characteristic such example is the phenomenon of plasticity of crystalline and amorphous materials that strongly depends on the material history. In plasticity modeling, the history is captured by a quenched, local and disordered flow stress distribution. While it is this disorder that causes avalanches that are commonly observed during nanoscale plastic deformation, the functional form and scaling properties have remained elusive.  In this paper,  a generic formalism is developed for deriving local disorder distributions from field-response (\eg stress/strain) timeseries in models of crackling noise. We demonstrate the efficiency of the method in the hysteretic random-field Ising model and also, models of elastic interface depinning that have been used to model crystalline and amorphous plasticity. We show that the capacity to resolve the quenched disorder distribution improves with the temporal resolution and number of samples.
}
\end{abstract}

\maketitle 

When Gibbs proposed the famous measure for quantifying the probabilities of individual gas microstates~\cite{Georgii:2011fk}, the primary interest was on figuring out the microscopic principles that gave rise to well defined thermodynamic state variables, even though actual microstates kept changing through Brownian motion. Gibbs' original inquiry has re-emerged in the context of material deformation of crystals but also other mechanically deformable systems: the far-from-equilibrium response  to mechanical deformation depends critically, especially at non-self-averaging small scales, on the initial configuration of defects. { Here, we provide a way that can be used, in principle, to estimate the probability of quenched microstates as a function of material parameters (plastic strain, size, hardening coefficients)\cite{Truesdell:1960fk}.}

The evidence for quenched disorder in initial defect microstructures has been accumulated through observations of abrupt plastic events or material-crackling noise in a large variety of materials, such as FCC and BCC crystals~\cite{uchic2009plasticity,Papanikolaou:2017zl,greer2011plasticity,Papanikolaou:2012kl}, amorphous solids~\cite{Sethna:2016fv} and also earthquake geological faults~\cite{uhl2015universal,bak2002unified}. This crackling noise \cite{Sethna:2001qf} has been commonly explained by random field models~\cite{Kuntz:2000bh,Martens:2012zr} or interface depinning ones~\cite{Fisher:1998fk,talamali2011avalanches,Nattermann:1992vn,Marchetti:2000ve,zaiser2006scale,Laurson:2012pi}, where the major component is homogeneous solid elasticity, but also a spatially inhomogeneous and random distribution of local, quenched disorder (typically entering local flow stress information)~\cite{Papanikolaou:2017zl,zaiser2006scale,ovaska2015quenched,Ispanovity:2014ve,Ispanovity:2013dq} and the allowed microstates are characterized by its stress and strain and minimize the elastic energy functional. { The evidence of crackling noise has led to an extensive study of the local, statistical properties of abrupt events and their properties, such as their sizes, durations, average shapes and their critical exponents. However, the major concern has been the fact that while homogeneous elastic properties are relatively straightforward to measure and test at virtually any scale~\cite{Asaro:2006fr}, the model distribution of local, quenched disorder is elusive, despite its commonly observed signature response of stochastic plastic bursts~\cite{uchic2009plasticity,Papanikolaou:2017zl,greer2011plasticity,Sethna:2016fv}. In this paper, we propose a feasible approach to ``learn" the quenched disorder distributions directly from load-response timeseries: We argue that the full characteristics of the timeseries may unveil the information on the form of the quenched disorder distribution which is not available through typical temporally local observables (such as abrupt event size/duration)~\cite{Papanikolaou:2011fu}. } While the major motivation of this work stems from plastic deformation, this method is generally applicable across crackling noise phenomena, defined through timeseries of an applied field (magnetic field, force, stress)  and the associated response variable (magnetization, displacement, strain). We label this method as Time Series - Machine Learning~(TS-ML). TS-ML is clearly limited by the physical applicability and completeness of the utilized model. { In its current form, for clarity purposes, TS-ML makes use of an unsupervised machine learning approach through principal component analysis (PCA) and k-means clustering. However, other unsupervised ML approaches could be successfully applied (see Supplementary Information for some examples).}   For demonstration purposes, TS-ML is applied to a typical model of plasticity~\cite{talamali2011avalanches}, as well as to a general example of crackling noise~\cite{Sethna:1993pd}. 

\begin{figure}[tbh]
\includegraphics[width=\textwidth]{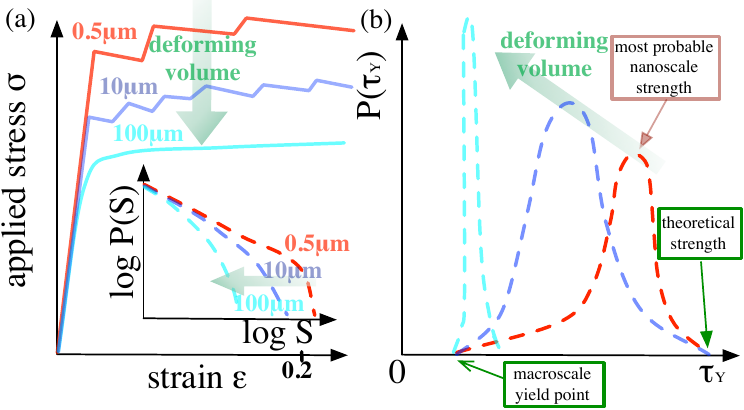}
\caption{{\bf {  Possible s}cenario of size effects and stochasticity in uniaxial compression of crystalline nano and micro pillars.} 
(a) Uniaxial stress-strain curves have been shown to be increasingly stochastic --with clear bursts-- and with higher apparent strength, as the sample/probe volume {\it decreases}, { in the range where the probed volume has effective diameter $0.5, 10,$ or $100\mu$m.~\cite{uchic2009plasticity,greer2011plasticity,Sethna:2016fv,Papanikolaou:2017zl}. The yield stress in such samples ranges from 50 to 500MPa. (inset):} Avalanche bursts, quantified through their strain magnitude $S$, have been shown to follow power-law distributions with a cutoff that decreases with sample volume.\cite{uchic2009plasticity,greer2011plasticity}.
(b) Increased strength and stochasticity at nanopillars could possibly originate into a nanoscale quenched yield distribution with { non-trivial wide form}, where the ``most probable" yield is displaced from the bulk yield point. This distribution should evolve into a normal distribution as sample volume increases, according to the central limit theorem. The very existence of  quenched disorder manifests in stochastic events that are common to describe through avalanches~\cite{Papanikolaou:2017zl}. { The green arrows display the direction of increasing representative volume that is being deformed.} 
}
\label{fig:fig1}
\end{figure}

Materials at small scales display size effects in material properties such as the strength and hardening coefficients~\cite{uchic2004sample,Maass:2017yq,Nix:1998ul, Ispanovity:2013dq,maass2015slip,Tsekenis:2013rr,Dahmen:2011ys,Maass:2017yq}. Stochastic yielding has been known to vanish with increasing system size~\cite{csikor2007dislocation}. However, its mechanism has been debated~\cite{Papanikolaou:2017zl,Maass:2017yq}. A plausible phenomenological scenario that explains the overall behavior is that the stochastic yield distribution at a representative volume has { non-trivially large tails}~\cite{Foss:2011rz} at the nanoscale~\footnote{that could either be a fat-tailed/large-kurtosis distribution or special distributions such as Weibull or Gumbel} and it gradually self-averages into the central limit behavior as the volume increases~(\cf Fig.~\ref{fig:fig1}). While still a phenomenologically-driven hypothesis, the validity of this scenario would explain the self-consistent emergence of size effects and stochasticity in small-scale plastic deformation,  in a material and sample independent manner~\cite{greer2005size,greer2011plasticity, Papanikolaou:2017zl}. { In contrast to this picture, it is common in multiscale modeling of material deformation~\cite{Asaro:2006fr, Lebensohn:1994zl} to assume that the statistics of yield parameters in a representative material volume are  dominated solely by their averages. Such a severe assumption is clearly not valid  in a scenario where the yielding distribution has a large variance, or it is non-Gaussian in nature. Given that all current multiscale modeling approaches are fundamentally based on estimates of atomistic mechanisms that are active at the nanoscale, it is natural to consider and pursue the validation of such a scenario. For this reason, robust and swift methods are required for the derivation of local yield distributions as function of sample size, prior loading history and time.
}

In this paper, we propose a systematic way to solve the inverse problem of deriving the stochastic yield distributions from timeseries that may be the outcome of mechanical deformation or other crackling noise experiments.  We use typical methods of unsupervised machine learning that naturally depend on physical modeling's completeness and the possible universality class~\cite{Papanikolaou:2017zl,Papanikolaou:2011fu}. The method does not aim to identify novel physical mechanisms of crackling noise; instead, it may provide a classification of parent quenched disorder distributions despite the existence of coexisting universal behavior. In the following, we perform simulations in two characteristic models of crackling noise, the $T=0$ mean-field Random Field Ising model~\cite{Sethna:1993pd}(RFIM) {  and the elastic long-range dipolar interface depinning model(LRDIDM)~\cite{Talamali:2011ij,zaiser2005fluctuation}}, designed for crystalline or amorphous solids. In all these cases, we observe a capacity of the method to cluster model data despite the similarity of avalanche distributions, and then classify them according to the quenched yield distribution at a testing volume. { Finally, the method may be also directly applied to actual experimental data on uniaxially compressed Ni micropillars, following prior work~\cite{Papanikolaou:2012kl} (see Section 4 and associated Fig.~S6 of the Supplementary Information).}
 \vspace{0.5in}

\begin{figure}[t!]
\includegraphics[width=\textwidth]{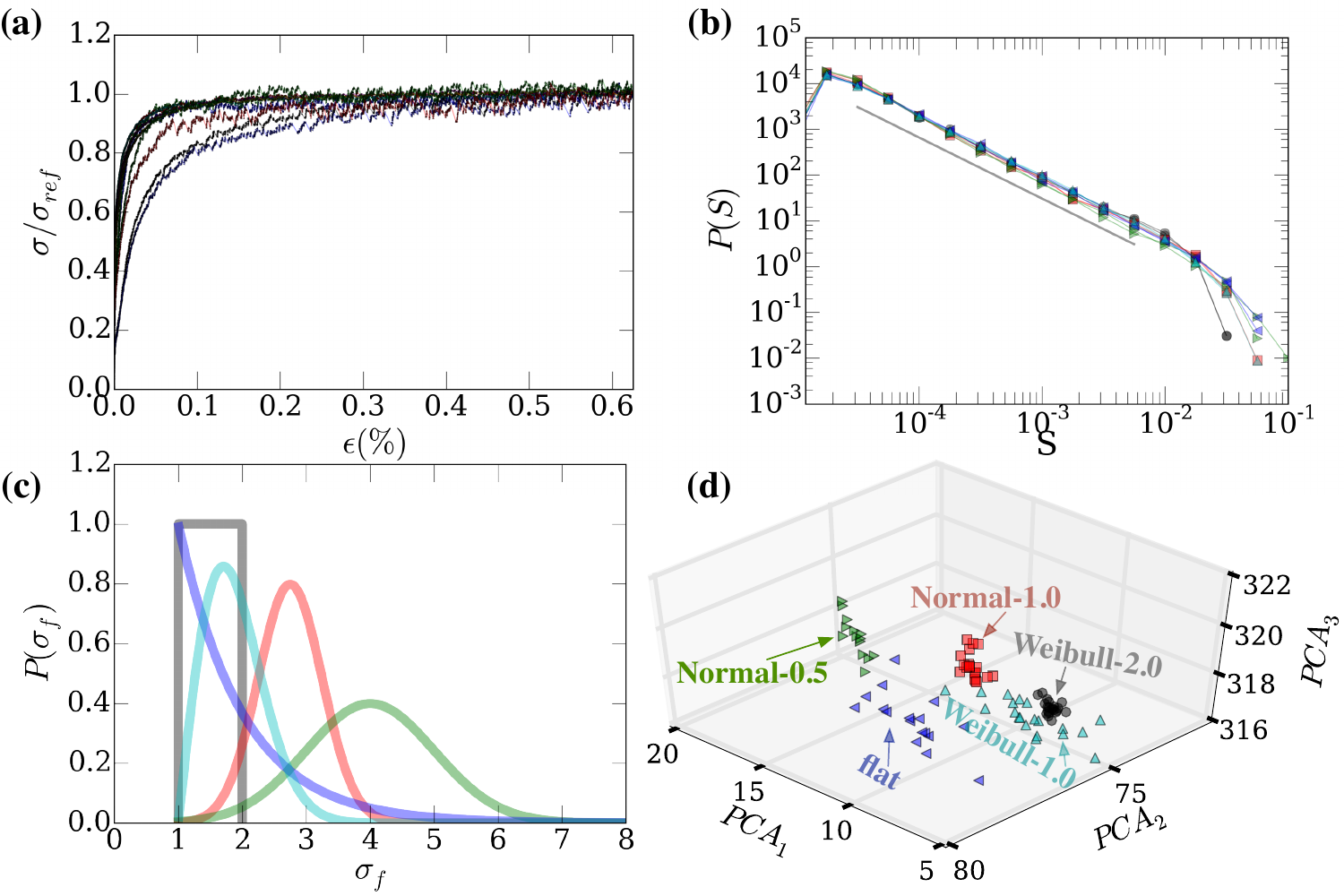}
\caption{
{\bf Learning the Input Quenched Flow Stress Distribution From Stress-Strain curves { in a model of crystal plasticity}.}
(a) Sample stress-strain curves for the continuous {  long-range dipolar  interface depinning model (LRDIDM),} originating in various local yield distributions (shown $2$ for each distribution). The stress is normalized by the stress value at $0.6\%$-strain. (b) The strain crackling event histograms $P(S)$ display an almost invariant power-law distribution with a large-size cutoff that is weakly influenced by the choice of quenched flow stress distributions. Disks correspond to the flat distribution (see (c)), squares to a normal distribution with variance 0.5, the right-pointing triangles correspond to a normal distribution with variance 1.0, the left-pointing triangles correspond to a Weibull distribution with $\alpha=1.0$ and the  upper-pointing triangles to $\alpha=2.0$. { The visible straight line segment is a power-law line guide to the eye with $y\sim x^{-1.35}$, as it is expected by theory. $S$ is measured in terms of strain units, so it is dimensionless. (c) The yield threshold distributions chosen in the model (where the model values for $\sigma_f$ are in units of $G/(2\pi (1-\nu))$ for the studied single-slip dislocation model studied~\cite{Papanikolaou:2017zl}), {  with the colors follow the descriptions in (b) and (d)}. } (d) Projection of samples on the 3D PCA space. Clear clustering is observed and the correspondence to the variety of distributions is shown. Clustering improves with number of samples, only 100 total samples were used for this example, with three decades of power-law events. { The symbol types match the ones shown in (b), in terms of the distribution being represented. Validation of the method for this model is discussed in the Supplementary Information, Section 2, Fig.S3.} 
}
\label{fig:fig4}
\end{figure}
{\bf Results}\\
TS-ML assumes $N$ samples that generally would include the discrete timeseries of the applied field $f_{ij}$ and the material response $m_{ij}$ with time index $i\in[1,T]$ and sample index $j\in[1,N]$. The complete data matrix is built through a multiscale description of the timeseries that may be performed through sequential estimation of the sliding window time-varying moments of order $n$, at scale $p$ and for sample $j$~\cite{Scoville:2015fk}
\bea
\langle X^{(n)}\rangle_{pj} = \sum_{k=0}^{p-1} \sum_{i=k 2^{-p} T+1}^{(k+1)2^{- p}T} ( X_{ij} - \langle X\rangle_{kj})^n 
\eea
where $X$ could be either $m$ or/and $f$. Through the construction of these moments, all fluctuations are captured (that could be equivalently unraveled through histogram distributions (\cf Fig.~\ref{fig:fig1}(b))). For each sample $j$, a vector may be constructed that contains all moments up to a maximum { resolution} scale $p_{\rm max}$ and max order $n_{\rm max}$, with total length $M=n_{\rm max}p_{\rm max}$. { The parameters  $n_{\rm max}$ and $p_{\rm max}$ are controlled by the resolution of the timeseries and can be estimated through: $p_{\rm max}\simeq log_2(T)/2$ and $n_{\rm max}\in [3, 8]$, given that identified moments remain non-zero for the given resolution.} Then, the effective data matrix $D_{\rm eff}$ is built through these vectors and  has dimensions $N\times M$. $D_{\rm eff}$ is used towards unsupervised machine learning through principal component analysis (PCA) and k-means clustering.

\begin{figure}[htb]
\includegraphics[width=\textwidth]{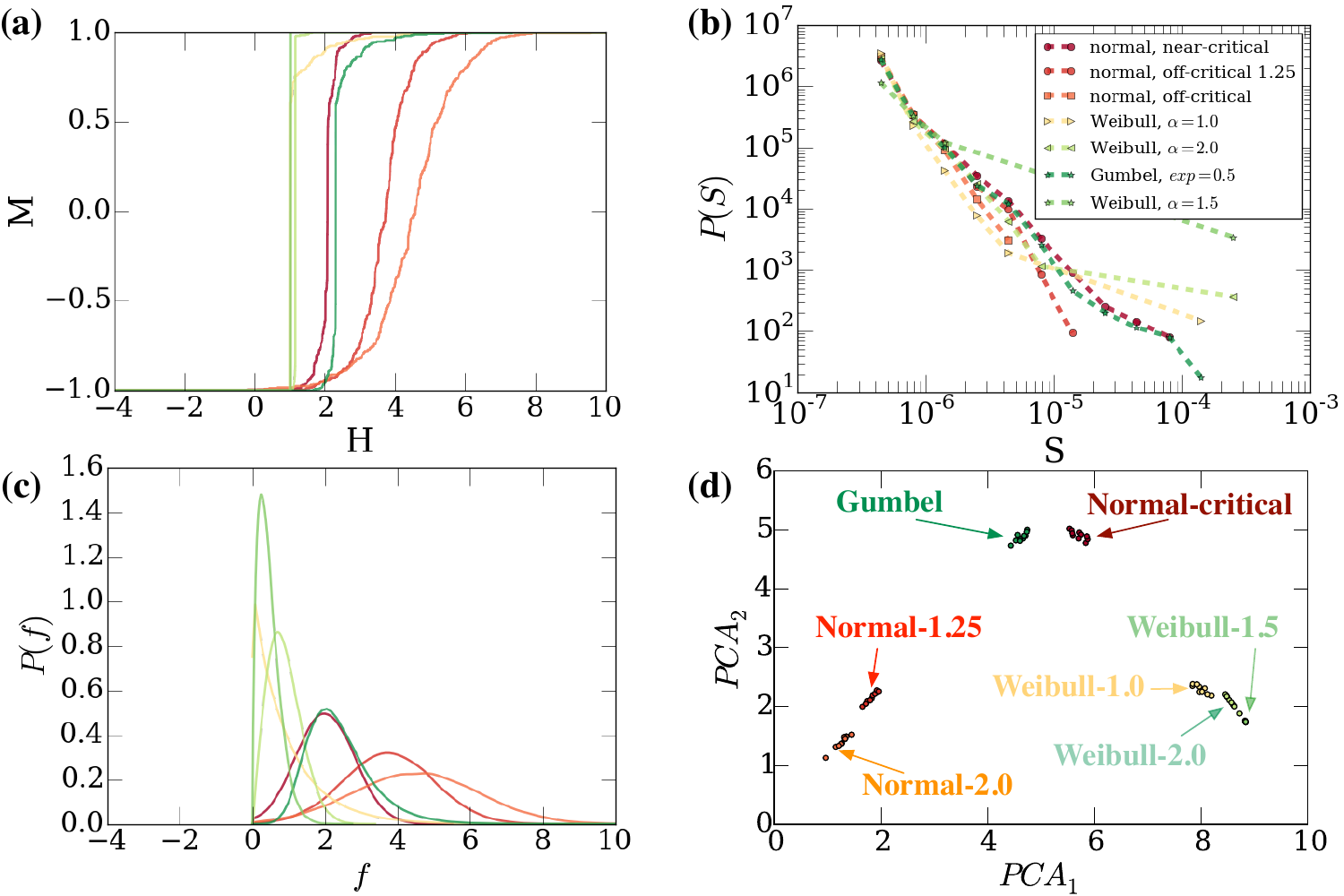}
\caption{{\bf Learning the Input Random Field Distribution Through Avalanche Data { in the RFIM}.}
(a) Sample $M-H$ timeseries. { Magnetization per spin $M$ ranges from -1 to +1, while the magnetic field $H$'s units are in terms of the spin-spin Ising interaction strength~\cite{Sethna:1993pd}.} Different colors correspond to different random distributions, from lighter to darker, we have the distributions of the { Upper Right legend (and the associated symbols shown in the legend)}: (1) Normal distribution with $\sigma=0.868\mu$, (2) Normal with $\sigma=1.25\mu$, (3) Normal distribution with $\sigma=2.0\mu$, (4) Weibull distribution with $\alpha=1.0$, (5) Weibull distribution with $\alpha=2.0$, (6) Gumbel distribution with exponent 0.5, (7) Weibull distribution with $\alpha=1.5$. 
(b) The magnetization crackling event distributions $P(S)$ are shown, (c) The randomness probability distributions for all 7 distributions, as they are input, (d) The projections of 140 samples (20 for each distribution) on the first 2 PCA components, according to the proposed method. Consistent clustering is observed { in the space spanned by only two PCA components. Clearly, the clustering is visible in the three dimensional PCA space (not shown) as in Fig.~\ref{fig:fig4}. { Symbol types and colors match the ones shown in (b) and the associated legend.} Validation of the method for this model is discussed in the Supplementary Information, Section 2, Fig.S3.}
}
\label{fig:fig2}
\end{figure}

Machine learning (ML) has had important success in various fields of science and engineering, but also materials science, for example accurate predictions of phase diagrams, crystal structures, and materials properties~\cite{Mueller:2016cr}. Unsupervised  learning through using PCA may work independentently of the input data types, as soon as the data cluster into distinct, distinguishable spaces that can be tracked by a clustering technique such as k-means~\cite{Press:1987oq}.  PCA reduces the dimensionality of a dataset~\cite{Greenewald:2015nx} by identifying subspaces that demonstrate characteristic data variation. The success of the method depends on the capacity of data that originate in different quenched disorder distributions, to be spatially separated into distinct clusters in PCA space. The identification and quantification of the clusters is done through the k-means method, which finds $k$ clusters that minimize the pairwise squared deviations of points in the same cluster. 
It is worth noting that our utilized ML is just a general way to statistically distinguish different signals, where the differences are caused by quenched disorder. However, other ML methods, may be similarly applicable (see Supplementary Information, Section 1 and Fig.S1), or may be expected to perform much better, such as deep learning methods~\cite{dl1,dl2}. These approaches will be studied in detail elsewhere.

{\it The models -- } TS-ML is applied on two characteristic models of crackling noise: { a) the elastic {  long-range dipolar interface-depinning model~\cite{Fisher:1998fk} (LRDIDM)} and b) the hysteretic Random-Field Ising Model (RFIM)~\cite{Sethna:1993pd}. The interactions in both models can be short or long ranged, altering the universality class. The purpose of this paper is to demonstrate examples, so some limiting behaviors are considered: mean-field interactions in the {  LRDIDM}~\cite{Dahmen:2009kl} and a complex, anisotropic and long-range kernel for the RFIM~\cite{Talamali:2011ij}. The method is identically applicable for quite general interaction kernels in both models. In the {  LRDIDM}, the Talamali \etal's approach~\cite{Talamali:2011ij} in two dimensions and assume that plastic deformation in disordered solids is modeled by the $xx-$component of the strain tensor $\eps^{(p)}\equiv\eps^{(p)}_{xx}=-\eps^{(p)}_{yy}$. The interaction due to local slip is the stress generated by local deformations of a random medium~\cite{Eshelby:1957fv}, which takes the form 
$\tilde F_{\rm int}(k,\omega)=(-\cos(4\omega)-1)\tilde \eps(k,\omega)$ 
where $k,\;\omega$ are the polar coordinates in Fourier space~\cite{Budrikis:2013fk} and $\tilde\eps, \;\tilde F_{\rm int}$ are the transforms of strain and interaction respectively. We initialize the $L\times L$ system with $\eps(\bfx)=0\;\forall \;\bfx$ and stress thresholds $f_p(\bfx)$ are taken from a uniform distribution $[0,1)$. The external field $F$  is increased adiabatically, and at each time-step, for all $\bfx$ that the total local force $f(\bfx)=F_{\rm int}(\bfx) + F - f_p(\bfx)$ becomes positive, there is a strain increase $d\eps(\bfx)$ randomly picked from a uniform $[0,1)$ distribution.
The external stress is decreased by $k/L^2$ at each time-step to cut-off  avalanches -- this cutoff mechanism resembles typical machine response at slow nominal straining.~\cite{Talamali:2011ij,Papanikolaou:2016rt}. In the RFIM, the approach follows the basic algorithm~\cite{Kuntz:1998vl,Kuntz:1999zp}, for which the zero temperature energy is ${\cal H}=-\sum_{i}(H+J+h_i)s_i$, with $H$ the applied field, $J$ the homogeneneous mean-field coupling, and $h_i$ the quenched random field that follows the distribution $\rho(h)$ (to be specified). The local spin $s_i$ gives rise to the response $m=\sum_{i=1}^N{s_i}$. Starting with $s_i=-1\forall i$, $H$ is increased until $s_i=+1\forall i$. In both models, it is possible to define abrupt events in terms of the size of response $S$ (strain in the {  LRDIDM}, magnetization per site in RFIM) during an event, as well as other quantities such as duration and energy-release~\cite{Papanikolaou:2017zl}. The distribution of $S$ has been studied extensively~\cite{Sethna:1993pd,Talamali:2011ij,Fisher:1998fk} and it is known to be described, in the pinned regime, by $P(S)\sim S^{-\tau}{\cal P}(S/S_0)$ where $\tau$ is $\sim1.3$~\cite{Budrikis:2013fk} for the {  LRDIDM} and $1.5$ for RFIM~\cite{Sethna:1993pd}. The cut-off function ${\cal P}(x)$ typically resembles an exponential function and $S_0$ is a nominal maximum event size. In the depinned regime, there are system-spanning events, and therefore $P(S)$ may include an additive component $P_{\rm inf}$ that scales with the system size. In this paper, we consider various behavioral regimes.}

\begin{figure}[htb]
\includegraphics[width=\textwidth]{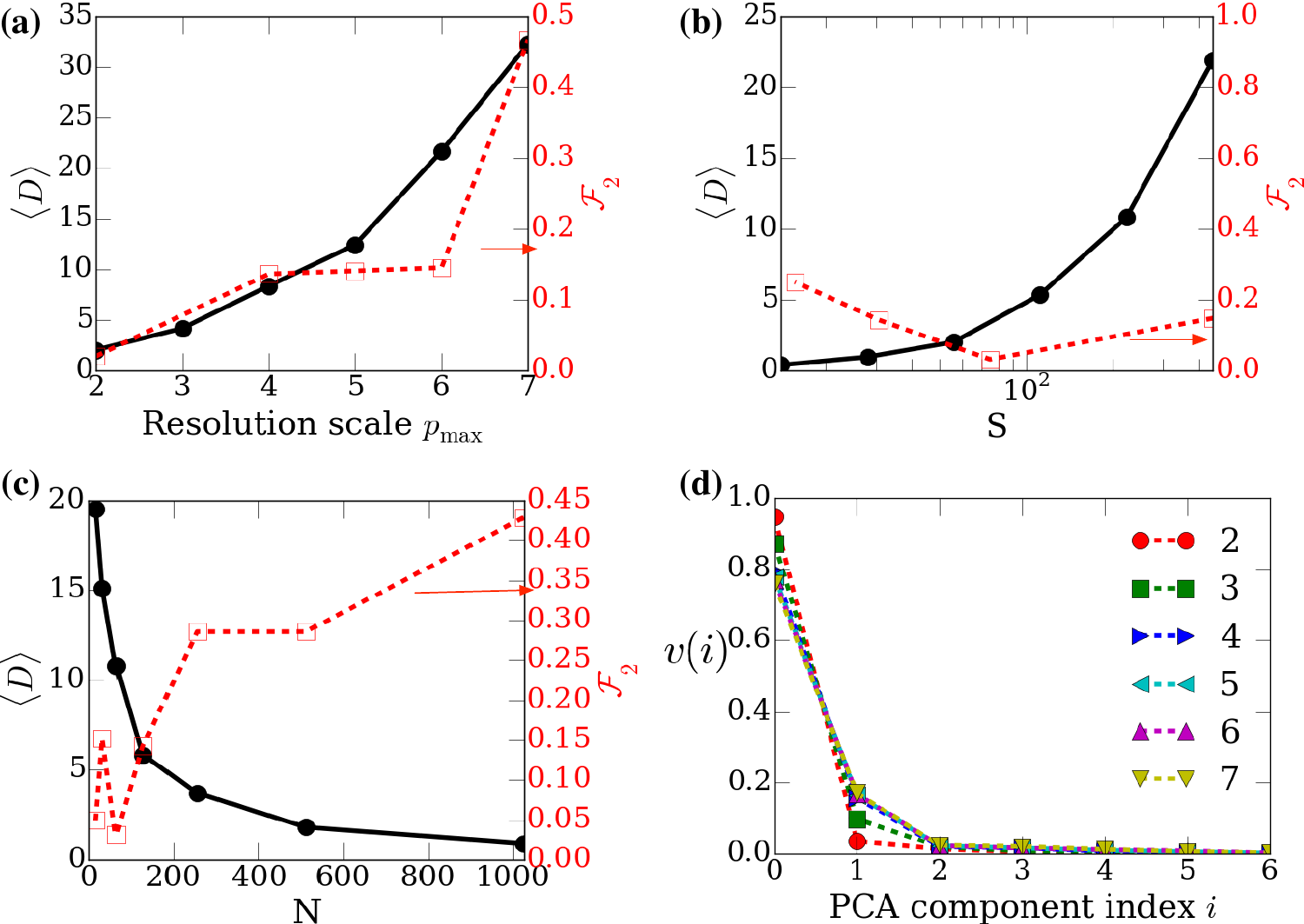}
\caption{
{\bf k-means Clustering Statistics \& Reliability for the mean-field Random Field model.}
{ (a) Inertia $\langle D\rangle$ (black-solid data) and $F_2$-score~\cite{f2score} (red-dashed) as function of the resolution scale parameter $p_{\rm max}$. (b) Inertia $\langle D\rangle$ and $F_2$-score as function of the number of samples S. (c) Inertia $\langle D\rangle$ and $F_2$-score as function of the number of system degrees of freedom. (d) Variance ratio $v(i)$ is shown in a sorted manner as function of the index $i$ of PCA components for different window resolution scale parameter $p_{\rm max}$. In (a) we choose $S=448$, $N=256$, and the resolution parameter $p_{\rm max}$ is varied. In (b), we choose $N=256$, $k=5$ and $S$ is varied. In (c) $S=448$, $k=5$ and $N$ is varied. 
}}
\label{fig:fig3}
\end{figure}

{\it Quenched Disorder Distributions -- } 
Various possible random distributions are considered in both models.   We consider: a) normal distributions with mean $\mu$ and variance $\sigma$, b)flat in a specified range, c)Weibull with probability density function ${g}(f)=\frac{\alpha}{\lambda} (f/\lambda)^{\alpha-1}e^{-(f/\lambda)^{\alpha}}$
 where $\delta f \sim N^{-1/\alpha}$ defines the range of the distribution. Here, $\alpha$ is the Weibull exponent \cite{Weibull:1951bh}, d) Gumbel distributions with ${g}(f)=e^{-(f - \mu)/\beta}e^{- e^{-(f-\mu)/\beta}}/\beta$ with mean $\mu+0.57721\beta$ and variance $\pi^2\beta^2/6$. All these distributions are very typical in materials science, depending on the character and origin of randomness~\cite{Bouchaud:1997tg}.

First, in order to demonstrate the applicability of TS-ML method in realistic {  plastic yield} problems, we consider the behavior of the {  LRDIDM}. As shown in Fig.~\ref{fig:fig4}, the input random distributions that range from Weibull, Normal and Flat (\cf Fig.~\ref{fig:fig4}(c)), give a range of nonlinear stress-strain curves (\cf Fig.~\ref{fig:fig4}(a)), consistent with the commonly observed behavior in crystalline and amorphous systems~\cite{Papanikolaou:2017zl}. However, abrupt event distributions in terms of the strain burst sizes, are characteristically independent of the imposed distribution (\cf Fig.~\ref{fig:fig4}(b)). This result is consistent with previous studies in various avalanche models~\cite{Le-Doussal:2009sp}. 

{ The presence of the abrupt event distribution validates TS-ML and facilitates a self-consistent physical picture. In order for TS-ML to be applicable (since high-order temporal moments need to be non-trivial), the presence of noise is required in the behavior. If such uncertainty was absent, then a simple constitutive formula would suffice to perform parameter fitting of the behavior. The very presence of abrupt events in the behavior validates the application of TS-ML, since it {  implies} that the overall behavior is driven by noise {  -- thus, a noise distribution should be identified}. In crystal plasticity, the presence of such {  abrupt} events further validate their origin into a quench disorder distribution, therefore TS-ML may be attempted on reasonable physical grounds. In such a noisy case (like, for example, in the experimentally testable case of uniaxially compressed Ni micropillars~\cite{Papanikolaou:2012kl}, studied in Section 4 of the Supplementary Information), it makes no physical or practical sense to apply parameter fitting for parameters such as work hardening or yielding -- {  these features are caused by a quenched disorder distribution that can be identified using TS-ML}. In a natural way, work hardening is caused by the form of the quenched disorder distribution, which can {  be} directly associated to the distribution of {\it e.g.} dislocation sources. If no such abrupt events {  are} observed, TS-ML {  should be regarded as unphysical/unnnatural, since there would be no motivation for a quenched disorder distribution in material modeling}. In this sense, the self-consistency of the physical picture may be tested. The application of TS-ML on 100 samples (20 each random distribution) is shown in Fig.~\ref{fig:fig4}, justifying the applicability of the method.
}

Then, the RFIM is discussed, which does not include any spatial resolution. In this model, as shown in Fig.~\ref{fig:fig2}, 20 samples of 7 different distributions are produced, (see Fig.~\ref{fig:fig2}(c)), totaling 140 samples. Then, the type of imposed randomness naturally influence the form of the $M-H$ curves (\cf Fig.~\ref{fig:fig2}(a)), however the avalanche distributions  (\cf Fig.~\ref{fig:fig2}(b)) demonstrate either a pre-depinning power-law behavior, or a spanning behavior, depending on the ratio of the disorder variance and the mean-field coupling strength. Then, all these samples are clustered and classified using TS-ML, and  a very clear clustering effect that captures completely the original quenched threshold distribution is seen (\cf Fig.~\ref{fig:fig2}(a)).

{
The behavior of the RFIM is analyzed in further detail, by considering the limits of the applicability of TS-ML (for the method validation, see also Supplementary Information, Sections 2 and 3 and associated Figs.~S3, S4 and S5). As shown in Fig.~\ref{fig:fig3}, the performance of the method is influenced as the number of samples $S$, the timeseries resolution $R$, and the number of degrees of freedom in the crackling system $N$. Three relevant observables are considered: 
\been
\item The ``inertia" of k-means clustering $\langle D\rangle$, which amounts to the sum of distances from the center of the assigned cluster; \item The $F_\beta$ score with $\beta=2$ which amounts to the success of the classification approach (k-means clustering) to identify correctly which samples are in the clusters~\cite{f2score}; The definition of the $F_2$ score  amounts to a weighted average of precision and recall: $F_2=$5$\times$precision$\times$recall/(4$\times$precision+recall), where Precision is defined as TP/(TP+FP), and Recall is defined as TP/(TP+FN). The $F_2$-score originates from the binary classification background, where we only have two classes that we want to distinguish are positive and negative. In this scenario there are four possible outcomes: TP (True Positive) $\rightarrow$ The object belongs to class positive and we classified it as positive, FP (False Positive ) $\rightarrow$ The object belongs to class negative and we classified it as positive, TN (True Negative) $\rightarrow$ The object belongs to class negative and we classified it as negative, FN (False Negative) $\rightarrow$ The object belongs to class positive but we classified it as negative. 
\item The variance percentage $v(i)$ of the PCA component $i$. As it can be seen, $\langle D\rangle$ increases with resolution $R$ (\cf Fig.~\ref{fig:fig3}(a)) and number of samples $S$ (\cf Fig.~\ref{fig:fig3}(b)), which amounts to naturally ``larger" clusters in the PCA space. However, $F_2$ decreases with both $R$, $S$ (\cf Fig.~\ref{fig:fig3}(a, b)), signifying that the larger clusters are also mutually well displaced, keeping the reliability of the method robust. Moreover, $\langle D \rangle$ and $F_2$ decrease drastically with the number of system degrees of freedom $N$. Finally, the variance is clearly distributed among the first $3$ components, with more than $97\%$ of the data variance. 
\een

}

\vspace{0.5cm}
{\bf Discussion}\\

{ 
In this paper, we {  suggested a possible scenario for size effects in crystal plasticity} and also, we proposed an explicit method for correlating the shape of noisy load-response curves in crackling noise phenomena to the shape of the quenched disorder distribution, which is the natural cause of the noise in the response timeseries. In the case of crystalline deformation of nanocrystals, we argued that this method and approach can investigate in detail the uncertainty in the work hardening behavior during statistical sampling of the deformation, in order to provide an estimate of the originating quenched disorder distribution that may relate to various physical quantities (e.g. precipitate defects or pre-existing immobile dislocation forests). This approach can be equivalently thought as an unbiased ``fitting" approach for work-hardening curves using stochastic distributions (instead of constitutive formulas that would only utilize average behaviors). In this paper, we argued that in various physical phenomena (especially ones that display crackling noise effects), the very nature of the problem enforces the use of stochastic distributions in order to be able to pursue multiscale modeling using physically accurate (statistically) microscopic information.

{  
The TS-ML method provides a unique description of the quenched disorder distribution in the limit of a large number of test samples, in the sense that the PCA clusters separate distinctly and can be easily distinguished by a classifier (\eg K-means). 
In the context of our theoretical models, we provide such evidence (\cf Fig. S3) in both the RFIM and the LRDIDM plasticity models (\cf Fig.4). Also, it is natural to check the cluster centers (identified through the centers of the K-Means clusters) and how they progress with the increase in the number of samples. As it is discussed in the Supplementary Information (\cf Fig. S4) the average cluster position for all clusters in the RFIM case (\cf Fig.~S2) is quite stable with the increase of the number of samples. Analogous results are found for the LRDIDM model. Moreover, uniqueness is critically dependent on the physical completeness of the utilized theoretical model. In that respect, the uniqueness of the TS-ML method can be established by the generality of its applicability: the {  LRDIDM} and RFIM models can be generalized by adding short or long range interactions, valid for various applications. We tested TS-ML for a number of additional possibilities, such as the 3D-RFIM hysteretic model~ \cite{Kuntz:1999zp}, the model of Ref.\cite{zaiser2005fluctuation} for continuum dislocation plasticity, and also discrete dislocation dynamics models of crystal plasticity~\cite{Papanikolaou:2015wt}. TS-ML appears to be generally applicable towards identifying quenched disorder distributions, appropriately defined for each model of reference. Finally, in continuum plasticity models the threshold distribution edge is critical for the yield behavior during avalanches. Thus, if two distributions have large overlap near the edge, then the limit of number of samples needed to distinguish different quenched disorder behaviors is prohibitively large (see SI's Sec.~3 and Fig.~S5 for a characteristic example, where overlapping Gaussian distributions lead to corresponding distinguishable but highly overlapping PCA maps). 
}

{ 
The validation of the method can be performed by separating data sets into ``experiment" and ``simulations", with the experimental data sets being the ones that will be assessed, while all the rest contributing to the classification scheme. As it is shown in Fig.~S3, we separated statistical samples into two equal subsets for both the RFIM and the LRDIDM plasticity model,  (``experimental"/testing and ``simulations"/training), and then the experimental data sets are projected on the trained PCA components. The testing data set successfully correlates in a percentage of around 90\% for this size of the data set (which includes 50-100 samples per quenched noise distribution). 
However, we acknowledge that the method requires further validation, especially in the experimental front: In the current work, we showed how the method may be practically applied to actual experimental data on uniaxially compressed Ni micropillars, following prior work~\cite{Papanikolaou:2012kl} (see Supplementary Information, Section 4 and  Fig.S6). Through this investigation, a quenched disorder distribution with Weibull statistics and $\alpha\simeq2$) was shown to be consistent with various theory-based suggestions, originating on Weibull statistics of dislocation sources (single-arm etc.) (see for example Ref.~\cite{Parthasarathy:2007fk}). Overall, the method  is a concrete, unsupervised classifier of noisy stress-strain curves (and possibly other timeseries) that takes into account all moments of the noise, beyond the average behavior.
}

Through this work, it has become clear that it is viable to distinguish random threshold distributions from stochastic field-response timeseries in typical crackling noise models. TS-ML can be efficiently implemented in order to distinguish quenched stochastic yield distributions in plastic deformation, that may originate in nanoscale experimental data.For this reason, high throughput experiments are required in order to efficiently probe the uncertainty in well defined deformed volumes. These distributions { help define the material properties, if a generalized definition of a ``material" is used where processing and deformation history is considered as a defining characteristic}. In the future, the target would be to classify these distributions for various experimental cases in the { processing} mechanics of various materials and also various prior loading histories, aiming at producing a library of stochastic yield distributions that can be implemented in multiscale mechanics models~\cite{Asaro:2006fr, Lebensohn:1994zl}, in a similar way that interatomic potentials libraries~\cite{Tadmor:2011ty} exist for molecular dynamics models~\cite{Frenkel:2002qv}. The crucial importance of such a physical picture is the transition to predictive models that are based on the self-consistent and intrinsically out-of-equilibrium statistical mechanics of crystal and amorphous plasticity.

}

\vspace{0.5cm}
{\bf Methods}\\
For the plasticity model, a custom Python code is used, solving the model on a $256\times 256$ grid. The samples were loaded on well equilibrated strain configurations, that were generated by repeated loading/unloading of the samples. $k$ is selected as 0.01 for these simulations. Typical cellular automaton rules are used, as described in the text. For the random-field model, the coupling $J$ is chosen to be unity and the number of spins is chosen to be $512$. For applying PCA, the sklearn library of Python is used: Non-linear PCA is accomplished by the application of Singular Value Decomposition, where the data matrix may be decomposed to a diagonal matrix of singular values $S$ and left/right singular vectors V/U, $D_{\rm eff}=V^T S U$. The $V$ vectors that correspond to the largest singular values, capture the most characteristic temporal behavior.  For solving the k-means problem, LloydÕs algorithm~\cite{Lloyd:1982hb} is used; the average complexity is given by $O(k n T)$, were $n$ is the number of samples and $T$ is the number of iteration, while the worst case complexity is given by $O(N^{k+2/M})$~\cite{Arthur:2006kl}. For more details, see also the Supplementary Information. {  The computational codes for the application of TS-ML on generic stress-strain or other timeseries are available upon e-mail request at {\rm stefanos.papanikolaou@mail.wvu.edu}. }


\begin{thebibliography}{59}%
\makeatletter
\providecommand \@ifxundefined [1]{%
 \@ifx{#1\undefined}
}%
\providecommand \@ifnum [1]{%
 \ifnum #1\expandafter \@firstoftwo
 \else \expandafter \@secondoftwo
 \fi
}%
\providecommand \@ifx [1]{%
 \ifx #1\expandafter \@firstoftwo
 \else \expandafter \@secondoftwo
 \fi
}%
\providecommand \natexlab [1]{#1}%
\providecommand \enquote  [1]{``#1''}%
\providecommand \bibnamefont  [1]{#1}%
\providecommand \bibfnamefont [1]{#1}%
\providecommand \citenamefont [1]{#1}%
\providecommand \href@noop [0]{\@secondoftwo}%
\providecommand \href [0]{\begingroup \@sanitize@url \@href}%
\providecommand \@href[1]{\@@startlink{#1}\@@href}%
\providecommand \@@href[1]{\endgroup#1\@@endlink}%
\providecommand \@sanitize@url [0]{\catcode `\\12\catcode `\$12\catcode
  `\&12\catcode `\#12\catcode `\^12\catcode `\_12\catcode `\%12\relax}%
\providecommand \@@startlink[1]{}%
\providecommand \@@endlink[0]{}%
\providecommand \url  [0]{\begingroup\@sanitize@url \@url }%
\providecommand \@url [1]{\endgroup\@href {#1}{\urlprefix }}%
\providecommand \urlprefix  [0]{URL }%
\providecommand \Eprint [0]{\href }%
\providecommand \doibase [0]{http://dx.doi.org/}%
\providecommand \selectlanguage [0]{\@gobble}%
\providecommand \bibinfo  [0]{\@secondoftwo}%
\providecommand \bibfield  [0]{\@secondoftwo}%
\providecommand \translation [1]{[#1]}%
\providecommand \BibitemOpen [0]{}%
\providecommand \bibitemStop [0]{}%
\providecommand \bibitemNoStop [0]{.\EOS\space}%
\providecommand \EOS [0]{\spacefactor3000\relax}%
\providecommand \BibitemShut  [1]{\csname bibitem#1\endcsname}%
\let\auto@bib@innerbib\@empty
\bibitem [{\citenamefont {Georgii}(2011)}]{Georgii:2011fk}%
  \BibitemOpen
  \bibfield  {author} {\bibinfo {author} {\bibfnamefont {H.-O.}\ \bibnamefont
  {Georgii}},\ }\href@noop {} {\emph {\bibinfo {title} {Gibbs measures and
  phase transitions}}},\ Vol.~\bibinfo {volume} {9}\ (\bibinfo  {publisher}
  {Walter de Gruyter},\ \bibinfo {year} {2011})\BibitemShut {NoStop}%
\bibitem [{\citenamefont {Truesdell}\ and\ \citenamefont
  {Toupin}(1960)}]{Truesdell:1960fk}%
  \BibitemOpen
  \bibfield  {author} {\bibinfo {author} {\bibfnamefont {C.}~\bibnamefont
  {Truesdell}}\ and\ \bibinfo {author} {\bibfnamefont {R.}~\bibnamefont
  {Toupin}},\ }\enquote {\bibinfo {title} {The classical field theories},}\ in\
  \href@noop {} {\emph {\bibinfo {booktitle} {Principles of Classical Mechanics
  and Field Theory/Prinzipien der Klassischen Mechanik und Feldtheorie}}}\
  (\bibinfo  {publisher} {Springer},\ \bibinfo {year} {1960})\ pp.\ \bibinfo
  {pages} {226--858}\BibitemShut {NoStop}%
\bibitem [{\citenamefont {Uchic}\ \emph {et~al.}(2009)\citenamefont {Uchic},
  \citenamefont {Shade},\ and\ \citenamefont {Dimiduk}}]{uchic2009plasticity}%
  \BibitemOpen
  \bibfield  {author} {\bibinfo {author} {\bibfnamefont {M.~D.}\ \bibnamefont
  {Uchic}}, \bibinfo {author} {\bibfnamefont {P.~A.}\ \bibnamefont {Shade}}, \
  and\ \bibinfo {author} {\bibfnamefont {D.~M.}\ \bibnamefont {Dimiduk}},\
  }\href@noop {} {\bibfield  {journal} {\bibinfo  {journal} {Annual Review of
  Materials Research}\ }\textbf {\bibinfo {volume} {39}},\ \bibinfo {pages}
  {361} (\bibinfo {year} {2009})}\BibitemShut {NoStop}%
\bibitem [{\citenamefont {Papanikolaou}\ \emph {et~al.}(2017)\citenamefont
  {Papanikolaou}, \citenamefont {Cui},\ and\ \citenamefont
  {Ghoniem}}]{Papanikolaou:2017zl}%
  \BibitemOpen
  \bibfield  {author} {\bibinfo {author} {\bibfnamefont {S.}~\bibnamefont
  {Papanikolaou}}, \bibinfo {author} {\bibfnamefont {Y.}~\bibnamefont {Cui}}, \
  and\ \bibinfo {author} {\bibfnamefont {N.}~\bibnamefont {Ghoniem}},\
  }\href@noop {} {\bibfield  {journal} {\bibinfo  {journal} {arXiv preprint
  arXiv:1705.06843}\ } (\bibinfo {year} {2017})}\BibitemShut {NoStop}%
\bibitem [{\citenamefont {Greer}\ and\ \citenamefont
  {De~Hosson}(2011)}]{greer2011plasticity}%
  \BibitemOpen
  \bibfield  {author} {\bibinfo {author} {\bibfnamefont {J.~R.}\ \bibnamefont
  {Greer}}\ and\ \bibinfo {author} {\bibfnamefont {J.~T.~M.}\ \bibnamefont
  {De~Hosson}},\ }\href@noop {} {\bibfield  {journal} {\bibinfo  {journal}
  {Progress in Materials Science}\ }\textbf {\bibinfo {volume} {56}},\ \bibinfo
  {pages} {654} (\bibinfo {year} {2011})}\BibitemShut {NoStop}%
\bibitem [{\citenamefont {Papanikolaou}\ \emph {et~al.}(2012)\citenamefont
  {Papanikolaou}, \citenamefont {Dimiduk}, \citenamefont {Choi}, \citenamefont
  {Sethna}, \citenamefont {Uchic}, \citenamefont {Woodward},\ and\
  \citenamefont {Zapperi}}]{Papanikolaou:2012kl}%
  \BibitemOpen
  \bibfield  {author} {\bibinfo {author} {\bibfnamefont {S.}~\bibnamefont
  {Papanikolaou}}, \bibinfo {author} {\bibfnamefont {D.~M.}\ \bibnamefont
  {Dimiduk}}, \bibinfo {author} {\bibfnamefont {W.}~\bibnamefont {Choi}},
  \bibinfo {author} {\bibfnamefont {J.~P.}\ \bibnamefont {Sethna}}, \bibinfo
  {author} {\bibfnamefont {M.~D.}\ \bibnamefont {Uchic}}, \bibinfo {author}
  {\bibfnamefont {C.~F.}\ \bibnamefont {Woodward}}, \ and\ \bibinfo {author}
  {\bibfnamefont {S.}~\bibnamefont {Zapperi}},\ }\href@noop {} {\bibfield
  {journal} {\bibinfo  {journal} {Nature}\ }\textbf {\bibinfo {volume} {490}},\
  \bibinfo {pages} {517} (\bibinfo {year} {2012})}\BibitemShut {NoStop}%
\bibitem [{\citenamefont {Sethna}\ \emph {et~al.}(2016)\citenamefont {Sethna},
  \citenamefont {Bierbaum}, \citenamefont {Dahmen}, \citenamefont {Goodrich},
  \citenamefont {Greer}, \citenamefont {Hayden}, \citenamefont {Kent-Dobias},
  \citenamefont {Lee}, \citenamefont {Liarte},\ and\ \citenamefont
  {Ni}}]{Sethna:2016fv}%
  \BibitemOpen
  \bibfield  {author} {\bibinfo {author} {\bibfnamefont {J.~P.}\ \bibnamefont
  {Sethna}}, \bibinfo {author} {\bibfnamefont {M.~K.}\ \bibnamefont
  {Bierbaum}}, \bibinfo {author} {\bibfnamefont {K.~A.}\ \bibnamefont
  {Dahmen}}, \bibinfo {author} {\bibfnamefont {C.~P.}\ \bibnamefont
  {Goodrich}}, \bibinfo {author} {\bibfnamefont {J.~R.}\ \bibnamefont {Greer}},
  \bibinfo {author} {\bibfnamefont {L.~X.}\ \bibnamefont {Hayden}}, \bibinfo
  {author} {\bibfnamefont {J.~P.}\ \bibnamefont {Kent-Dobias}}, \bibinfo
  {author} {\bibfnamefont {E.~D.}\ \bibnamefont {Lee}}, \bibinfo {author}
  {\bibfnamefont {D.~B.}\ \bibnamefont {Liarte}}, \ and\ \bibinfo {author}
  {\bibfnamefont {X.}~\bibnamefont {Ni}},\ }\href@noop {} {\bibfield  {journal}
  {\bibinfo  {journal} {arXiv preprint arXiv:1609.05838}\ } (\bibinfo {year}
  {2016})}\BibitemShut {NoStop}%
\bibitem [{\citenamefont {Uhl}\ \emph {et~al.}(2015)\citenamefont {Uhl},
  \citenamefont {Pathak}, \citenamefont {Schorlemmer}, \citenamefont {Liu},
  \citenamefont {Swindeman}, \citenamefont {Brinkman}, \citenamefont {LeBlanc},
  \citenamefont {Tsekenis}, \citenamefont {Friedman}, \citenamefont {Behringer}
  \emph {et~al.}}]{uhl2015universal}%
  \BibitemOpen
  \bibfield  {author} {\bibinfo {author} {\bibfnamefont {J.~T.}\ \bibnamefont
  {Uhl}}, \bibinfo {author} {\bibfnamefont {S.}~\bibnamefont {Pathak}},
  \bibinfo {author} {\bibfnamefont {D.}~\bibnamefont {Schorlemmer}}, \bibinfo
  {author} {\bibfnamefont {X.}~\bibnamefont {Liu}}, \bibinfo {author}
  {\bibfnamefont {R.}~\bibnamefont {Swindeman}}, \bibinfo {author}
  {\bibfnamefont {B.~A.}\ \bibnamefont {Brinkman}}, \bibinfo {author}
  {\bibfnamefont {M.}~\bibnamefont {LeBlanc}}, \bibinfo {author} {\bibfnamefont
  {G.}~\bibnamefont {Tsekenis}}, \bibinfo {author} {\bibfnamefont
  {N.}~\bibnamefont {Friedman}}, \bibinfo {author} {\bibfnamefont
  {R.}~\bibnamefont {Behringer}},  \emph {et~al.},\ }\href@noop {} {\bibfield
  {journal} {\bibinfo  {journal} {Sci. Rep.}\ }\textbf {\bibinfo {volume}
  {5}},\ \bibinfo {pages} {16493} (\bibinfo {year} {2015})}\BibitemShut
  {NoStop}%
\bibitem [{\citenamefont {Bak}\ \emph {et~al.}(2002)\citenamefont {Bak},
  \citenamefont {Christensen}, \citenamefont {Danon},\ and\ \citenamefont
  {Scanlon}}]{bak2002unified}%
  \BibitemOpen
  \bibfield  {author} {\bibinfo {author} {\bibfnamefont {P.}~\bibnamefont
  {Bak}}, \bibinfo {author} {\bibfnamefont {K.}~\bibnamefont {Christensen}},
  \bibinfo {author} {\bibfnamefont {L.}~\bibnamefont {Danon}}, \ and\ \bibinfo
  {author} {\bibfnamefont {T.}~\bibnamefont {Scanlon}},\ }\href@noop {}
  {\bibfield  {journal} {\bibinfo  {journal} {Physical Review Letters}\
  }\textbf {\bibinfo {volume} {88}},\ \bibinfo {pages} {178501} (\bibinfo
  {year} {2002})}\BibitemShut {NoStop}%
\bibitem [{\citenamefont {Sethna}\ \emph {et~al.}(2001)\citenamefont {Sethna},
  \citenamefont {Dahmen},\ and\ \citenamefont {Myers}}]{Sethna:2001qf}%
  \BibitemOpen
  \bibfield  {author} {\bibinfo {author} {\bibfnamefont {J.~P.}\ \bibnamefont
  {Sethna}}, \bibinfo {author} {\bibfnamefont {K.~A.}\ \bibnamefont {Dahmen}},
  \ and\ \bibinfo {author} {\bibfnamefont {C.~R.}\ \bibnamefont {Myers}},\
  }\href@noop {} {\bibfield  {journal} {\bibinfo  {journal} {Nature}\ }\textbf
  {\bibinfo {volume} {410}},\ \bibinfo {pages} {242} (\bibinfo {year}
  {2001})}\BibitemShut {NoStop}%
\bibitem [{\citenamefont {Kuntz}\ and\ \citenamefont
  {Sethna}(2000)}]{Kuntz:2000bh}%
  \BibitemOpen
  \bibfield  {author} {\bibinfo {author} {\bibfnamefont {M.~C.}\ \bibnamefont
  {Kuntz}}\ and\ \bibinfo {author} {\bibfnamefont {J.~P.}\ \bibnamefont
  {Sethna}},\ }\href@noop {} {\bibfield  {journal} {\bibinfo  {journal}
  {Physical Review B}\ }\textbf {\bibinfo {volume} {62}},\ \bibinfo {pages}
  {11699} (\bibinfo {year} {2000})}\BibitemShut {NoStop}%
\bibitem [{\citenamefont {Martens}\ \emph {et~al.}(2012)\citenamefont
  {Martens}, \citenamefont {Bocquet},\ and\ \citenamefont
  {Barrat}}]{Martens:2012zr}%
  \BibitemOpen
  \bibfield  {author} {\bibinfo {author} {\bibfnamefont {K.}~\bibnamefont
  {Martens}}, \bibinfo {author} {\bibfnamefont {L.}~\bibnamefont {Bocquet}}, \
  and\ \bibinfo {author} {\bibfnamefont {J.-L.}\ \bibnamefont {Barrat}},\
  }\href {\doibase 10.1039/C2SM07090A} {\bibfield  {journal} {\bibinfo
  {journal} {Soft Matter}\ }\textbf {\bibinfo {volume} {8}},\ \bibinfo {pages}
  {4197} (\bibinfo {year} {2012})}\BibitemShut {NoStop}%
\bibitem [{\citenamefont {Fisher}(1998)}]{Fisher:1998fk}%
  \BibitemOpen
  \bibfield  {author} {\bibinfo {author} {\bibfnamefont {D.~S.}\ \bibnamefont
  {Fisher}},\ }\href@noop {} {\bibfield  {journal} {\bibinfo  {journal}
  {Physics reports}\ }\textbf {\bibinfo {volume} {301}},\ \bibinfo {pages}
  {113} (\bibinfo {year} {1998})}\BibitemShut {NoStop}%
\bibitem [{\citenamefont {Talamali}\ \emph
  {et~al.}(2011{\natexlab{a}})\citenamefont {Talamali}, \citenamefont
  {Pet{\"a}j{\"a}}, \citenamefont {Vandembroucq},\ and\ \citenamefont
  {Roux}}]{talamali2011avalanches}%
  \BibitemOpen
  \bibfield  {author} {\bibinfo {author} {\bibfnamefont {M.}~\bibnamefont
  {Talamali}}, \bibinfo {author} {\bibfnamefont {V.}~\bibnamefont
  {Pet{\"a}j{\"a}}}, \bibinfo {author} {\bibfnamefont {D.}~\bibnamefont
  {Vandembroucq}}, \ and\ \bibinfo {author} {\bibfnamefont {S.}~\bibnamefont
  {Roux}},\ }\href@noop {} {\bibfield  {journal} {\bibinfo  {journal} {Physical
  Review E}\ }\textbf {\bibinfo {volume} {84}},\ \bibinfo {pages} {016115}
  (\bibinfo {year} {2011}{\natexlab{a}})}\BibitemShut {NoStop}%
\bibitem [{\citenamefont {Nattermann}\ \emph {et~al.}(1992)\citenamefont
  {Nattermann}, \citenamefont {Stepanow}, \citenamefont {Tang},\ and\
  \citenamefont {Leschhorn}}]{Nattermann:1992vn}%
  \BibitemOpen
  \bibfield  {author} {\bibinfo {author} {\bibfnamefont {T.}~\bibnamefont
  {Nattermann}}, \bibinfo {author} {\bibfnamefont {S.}~\bibnamefont
  {Stepanow}}, \bibinfo {author} {\bibfnamefont {L.-H.}\ \bibnamefont {Tang}},
  \ and\ \bibinfo {author} {\bibfnamefont {H.}~\bibnamefont {Leschhorn}},\
  }\href@noop {} {\bibfield  {journal} {\bibinfo  {journal} {Journal de
  Physique II}\ }\textbf {\bibinfo {volume} {2}},\ \bibinfo {pages} {1483}
  (\bibinfo {year} {1992})}\BibitemShut {NoStop}%
\bibitem [{\citenamefont {Marchetti}\ \emph {et~al.}(2000)\citenamefont
  {Marchetti}, \citenamefont {Middleton},\ and\ \citenamefont
  {Prellberg}}]{Marchetti:2000ve}%
  \BibitemOpen
  \bibfield  {author} {\bibinfo {author} {\bibfnamefont {M.~C.}\ \bibnamefont
  {Marchetti}}, \bibinfo {author} {\bibfnamefont {A.~A.}\ \bibnamefont
  {Middleton}}, \ and\ \bibinfo {author} {\bibfnamefont {T.}~\bibnamefont
  {Prellberg}},\ }\href@noop {} {\bibfield  {journal} {\bibinfo  {journal}
  {Physical Review Letters}\ }\textbf {\bibinfo {volume} {85}},\ \bibinfo
  {pages} {1104} (\bibinfo {year} {2000})}\BibitemShut {NoStop}%
\bibitem [{\citenamefont {Zaiser}(2006)}]{zaiser2006scale}%
  \BibitemOpen
  \bibfield  {author} {\bibinfo {author} {\bibfnamefont {M.}~\bibnamefont
  {Zaiser}},\ }\href@noop {} {\bibfield  {journal} {\bibinfo  {journal} {Adv.
  Phys.}\ }\textbf {\bibinfo {volume} {55}},\ \bibinfo {pages} {185} (\bibinfo
  {year} {2006})}\BibitemShut {NoStop}%
\bibitem [{\citenamefont {Laurson}\ and\ \citenamefont
  {Alava}(2012)}]{Laurson:2012pi}%
  \BibitemOpen
  \bibfield  {author} {\bibinfo {author} {\bibfnamefont {L.}~\bibnamefont
  {Laurson}}\ and\ \bibinfo {author} {\bibfnamefont {M.~J.}\ \bibnamefont
  {Alava}},\ }\href@noop {} {\bibfield  {journal} {\bibinfo  {journal}
  {Physical review letters}\ }\textbf {\bibinfo {volume} {109}},\ \bibinfo
  {pages} {155504} (\bibinfo {year} {2012})}\BibitemShut {NoStop}%
\bibitem [{\citenamefont {Ovaska}\ \emph {et~al.}(2015)\citenamefont {Ovaska},
  \citenamefont {Laurson},\ and\ \citenamefont {Alava}}]{ovaska2015quenched}%
  \BibitemOpen
  \bibfield  {author} {\bibinfo {author} {\bibfnamefont {M.}~\bibnamefont
  {Ovaska}}, \bibinfo {author} {\bibfnamefont {L.}~\bibnamefont {Laurson}}, \
  and\ \bibinfo {author} {\bibfnamefont {M.~J.}\ \bibnamefont {Alava}},\
  }\href@noop {} {\bibfield  {journal} {\bibinfo  {journal} {Scientific
  reports}\ }\textbf {\bibinfo {volume} {5}} (\bibinfo {year}
  {2015})}\BibitemShut {NoStop}%
\bibitem [{\citenamefont {Isp{\'a}novity}\ \emph {et~al.}(2014)\citenamefont
  {Isp{\'a}novity}, \citenamefont {Laurson}, \citenamefont {Zaiser},
  \citenamefont {Groma}, \citenamefont {Zapperi},\ and\ \citenamefont
  {Alava}}]{Ispanovity:2014ve}%
  \BibitemOpen
  \bibfield  {author} {\bibinfo {author} {\bibfnamefont {P.~D.}\ \bibnamefont
  {Isp{\'a}novity}}, \bibinfo {author} {\bibfnamefont {L.}~\bibnamefont
  {Laurson}}, \bibinfo {author} {\bibfnamefont {M.}~\bibnamefont {Zaiser}},
  \bibinfo {author} {\bibfnamefont {I.}~\bibnamefont {Groma}}, \bibinfo
  {author} {\bibfnamefont {S.}~\bibnamefont {Zapperi}}, \ and\ \bibinfo
  {author} {\bibfnamefont {M.~J.}\ \bibnamefont {Alava}},\ }\href@noop {}
  {\bibfield  {journal} {\bibinfo  {journal} {Physical review letters}\
  }\textbf {\bibinfo {volume} {112}},\ \bibinfo {pages} {235501} (\bibinfo
  {year} {2014})}\BibitemShut {NoStop}%
\bibitem [{\citenamefont {Isp{\'a}novity}\ \emph {et~al.}(2013)\citenamefont
  {Isp{\'a}novity}, \citenamefont {Hegyi}, \citenamefont {Groma}, \citenamefont
  {Gy{\"o}rgyi}, \citenamefont {Ratter},\ and\ \citenamefont
  {Weygand}}]{Ispanovity:2013dq}%
  \BibitemOpen
  \bibfield  {author} {\bibinfo {author} {\bibfnamefont {P.~D.}\ \bibnamefont
  {Isp{\'a}novity}}, \bibinfo {author} {\bibfnamefont {{\'A}.}~\bibnamefont
  {Hegyi}}, \bibinfo {author} {\bibfnamefont {I.}~\bibnamefont {Groma}},
  \bibinfo {author} {\bibfnamefont {G.}~\bibnamefont {Gy{\"o}rgyi}}, \bibinfo
  {author} {\bibfnamefont {K.}~\bibnamefont {Ratter}}, \ and\ \bibinfo {author}
  {\bibfnamefont {D.}~\bibnamefont {Weygand}},\ }\href@noop {} {\bibfield
  {journal} {\bibinfo  {journal} {Acta Materialia}\ }\textbf {\bibinfo {volume}
  {61}},\ \bibinfo {pages} {6234} (\bibinfo {year} {2013})}\BibitemShut
  {NoStop}%
\bibitem [{\citenamefont {Asaro}\ and\ \citenamefont
  {Lubarda}(2006)}]{Asaro:2006fr}%
  \BibitemOpen
  \bibfield  {author} {\bibinfo {author} {\bibfnamefont {R.}~\bibnamefont
  {Asaro}}\ and\ \bibinfo {author} {\bibfnamefont {V.}~\bibnamefont
  {Lubarda}},\ }\href@noop {} {\emph {\bibinfo {title} {Mechanics of solids and
  materials}}}\ (\bibinfo  {publisher} {Cambridge University Press},\ \bibinfo
  {year} {2006})\BibitemShut {NoStop}%
\bibitem [{\citenamefont {Papanikolaou}\ \emph {et~al.}(2011)\citenamefont
  {Papanikolaou}, \citenamefont {Bohn}, \citenamefont {Sommer}, \citenamefont
  {Durin}, \citenamefont {Zapperi},\ and\ \citenamefont
  {Sethna}}]{Papanikolaou:2011fu}%
  \BibitemOpen
  \bibfield  {author} {\bibinfo {author} {\bibfnamefont {S.}~\bibnamefont
  {Papanikolaou}}, \bibinfo {author} {\bibfnamefont {F.}~\bibnamefont {Bohn}},
  \bibinfo {author} {\bibfnamefont {R.~L.}\ \bibnamefont {Sommer}}, \bibinfo
  {author} {\bibfnamefont {G.}~\bibnamefont {Durin}}, \bibinfo {author}
  {\bibfnamefont {S.}~\bibnamefont {Zapperi}}, \ and\ \bibinfo {author}
  {\bibfnamefont {J.~P.}\ \bibnamefont {Sethna}},\ }\href@noop {} {\bibfield
  {journal} {\bibinfo  {journal} {Nature Physics}\ }\textbf {\bibinfo {volume}
  {7}},\ \bibinfo {pages} {316} (\bibinfo {year} {2011})}\BibitemShut {NoStop}%
\bibitem [{\citenamefont {Sethna}\ \emph {et~al.}(1993)\citenamefont {Sethna},
  \citenamefont {Dahmen}, \citenamefont {Kartha}, \citenamefont {Krumhansl},
  \citenamefont {Roberts},\ and\ \citenamefont {Shore}}]{Sethna:1993pd}%
  \BibitemOpen
  \bibfield  {author} {\bibinfo {author} {\bibfnamefont {J.~P.}\ \bibnamefont
  {Sethna}}, \bibinfo {author} {\bibfnamefont {K.}~\bibnamefont {Dahmen}},
  \bibinfo {author} {\bibfnamefont {S.}~\bibnamefont {Kartha}}, \bibinfo
  {author} {\bibfnamefont {J.~A.}\ \bibnamefont {Krumhansl}}, \bibinfo {author}
  {\bibfnamefont {B.~W.}\ \bibnamefont {Roberts}}, \ and\ \bibinfo {author}
  {\bibfnamefont {J.~D.}\ \bibnamefont {Shore}},\ }\href@noop {} {\bibfield
  {journal} {\bibinfo  {journal} {Physical Review Letters}\ }\textbf {\bibinfo
  {volume} {70}},\ \bibinfo {pages} {3347} (\bibinfo {year}
  {1993})}\BibitemShut {NoStop}%
\bibitem [{\citenamefont {Uchic}\ \emph {et~al.}(2004)\citenamefont {Uchic},
  \citenamefont {Dimiduk}, \citenamefont {Florando},\ and\ \citenamefont
  {Nix}}]{uchic2004sample}%
  \BibitemOpen
  \bibfield  {author} {\bibinfo {author} {\bibfnamefont {M.~D.}\ \bibnamefont
  {Uchic}}, \bibinfo {author} {\bibfnamefont {D.~M.}\ \bibnamefont {Dimiduk}},
  \bibinfo {author} {\bibfnamefont {J.~N.}\ \bibnamefont {Florando}}, \ and\
  \bibinfo {author} {\bibfnamefont {W.~D.}\ \bibnamefont {Nix}},\ }\href@noop
  {} {\bibfield  {journal} {\bibinfo  {journal} {Science}\ }\textbf {\bibinfo
  {volume} {305}},\ \bibinfo {pages} {986} (\bibinfo {year}
  {2004})}\BibitemShut {NoStop}%
\bibitem [{\citenamefont {Maass}\ and\ \citenamefont
  {Derlet}(2017)}]{Maass:2017yq}%
  \BibitemOpen
  \bibfield  {author} {\bibinfo {author} {\bibfnamefont {R.}~\bibnamefont
  {Maass}}\ and\ \bibinfo {author} {\bibfnamefont {P.}~\bibnamefont {Derlet}},\
  }\href@noop {} {\bibfield  {journal} {\bibinfo  {journal} {arXiv preprint
  arXiv:1704.07297}\ } (\bibinfo {year} {2017})}\BibitemShut {NoStop}%
\bibitem [{\citenamefont {Nix}\ and\ \citenamefont {Gao}(1998)}]{Nix:1998ul}%
  \BibitemOpen
  \bibfield  {author} {\bibinfo {author} {\bibfnamefont {W.~D.}\ \bibnamefont
  {Nix}}\ and\ \bibinfo {author} {\bibfnamefont {H.}~\bibnamefont {Gao}},\
  }\href@noop {} {\bibfield  {journal} {\bibinfo  {journal} {Journal of the
  Mechanics and Physics of Solids}\ }\textbf {\bibinfo {volume} {46}},\
  \bibinfo {pages} {411} (\bibinfo {year} {1998})}\BibitemShut {NoStop}%
\bibitem [{\citenamefont {Maa{\ss}}\ \emph {et~al.}(2015)\citenamefont
  {Maa{\ss}}, \citenamefont {Wraith}, \citenamefont {Uhl}, \citenamefont
  {Greer},\ and\ \citenamefont {Dahmen}}]{maass2015slip}%
  \BibitemOpen
  \bibfield  {author} {\bibinfo {author} {\bibfnamefont {R.}~\bibnamefont
  {Maa{\ss}}}, \bibinfo {author} {\bibfnamefont {M.}~\bibnamefont {Wraith}},
  \bibinfo {author} {\bibfnamefont {J.}~\bibnamefont {Uhl}}, \bibinfo {author}
  {\bibfnamefont {J.}~\bibnamefont {Greer}}, \ and\ \bibinfo {author}
  {\bibfnamefont {K.}~\bibnamefont {Dahmen}},\ }\href@noop {} {\bibfield
  {journal} {\bibinfo  {journal} {Phys. Rev. E}\ }\textbf {\bibinfo {volume}
  {91}},\ \bibinfo {pages} {042403} (\bibinfo {year} {2015})}\BibitemShut
  {NoStop}%
\bibitem [{\citenamefont {Tsekenis}\ \emph {et~al.}(2013)\citenamefont
  {Tsekenis}, \citenamefont {Uhl}, \citenamefont {Goldenfeld},\ and\
  \citenamefont {Dahmen}}]{Tsekenis:2013rr}%
  \BibitemOpen
  \bibfield  {author} {\bibinfo {author} {\bibfnamefont {G.}~\bibnamefont
  {Tsekenis}}, \bibinfo {author} {\bibfnamefont {J.~T.}\ \bibnamefont {Uhl}},
  \bibinfo {author} {\bibfnamefont {N.}~\bibnamefont {Goldenfeld}}, \ and\
  \bibinfo {author} {\bibfnamefont {K.~A.}\ \bibnamefont {Dahmen}},\
  }\href@noop {} {\bibfield  {journal} {\bibinfo  {journal} {EPL (Europhysics
  Letters)}\ }\textbf {\bibinfo {volume} {101}},\ \bibinfo {pages} {36003}
  (\bibinfo {year} {2013})}\BibitemShut {NoStop}%
\bibitem [{\citenamefont {Dahmen}\ \emph {et~al.}(2011)\citenamefont {Dahmen},
  \citenamefont {Ben-Zion},\ and\ \citenamefont {Uhl}}]{Dahmen:2011ys}%
  \BibitemOpen
  \bibfield  {author} {\bibinfo {author} {\bibfnamefont {K.~A.}\ \bibnamefont
  {Dahmen}}, \bibinfo {author} {\bibfnamefont {Y.}~\bibnamefont {Ben-Zion}}, \
  and\ \bibinfo {author} {\bibfnamefont {J.~T.}\ \bibnamefont {Uhl}},\
  }\href@noop {} {\bibfield  {journal} {\bibinfo  {journal} {Nature Physics}\
  }\textbf {\bibinfo {volume} {7}},\ \bibinfo {pages} {554} (\bibinfo {year}
  {2011})}\BibitemShut {NoStop}%
\bibitem [{\citenamefont {Csikor}\ \emph {et~al.}(2007)\citenamefont {Csikor},
  \citenamefont {Motz}, \citenamefont {Weygand}, \citenamefont {Zaiser},\ and\
  \citenamefont {Zapperi}}]{csikor2007dislocation}%
  \BibitemOpen
  \bibfield  {author} {\bibinfo {author} {\bibfnamefont {F.~F.}\ \bibnamefont
  {Csikor}}, \bibinfo {author} {\bibfnamefont {C.}~\bibnamefont {Motz}},
  \bibinfo {author} {\bibfnamefont {D.}~\bibnamefont {Weygand}}, \bibinfo
  {author} {\bibfnamefont {M.}~\bibnamefont {Zaiser}}, \ and\ \bibinfo {author}
  {\bibfnamefont {S.}~\bibnamefont {Zapperi}},\ }\href@noop {} {\bibfield
  {journal} {\bibinfo  {journal} {Science}\ }\textbf {\bibinfo {volume}
  {318}},\ \bibinfo {pages} {251} (\bibinfo {year} {2007})}\BibitemShut
  {NoStop}%
\bibitem [{\citenamefont {Foss}\ \emph {et~al.}(2011)\citenamefont {Foss},
  \citenamefont {Korshunov},\ and\ \citenamefont {Zachary}}]{Foss:2011rz}%
  \BibitemOpen
  \bibfield  {author} {\bibinfo {author} {\bibfnamefont {S.}~\bibnamefont
  {Foss}}, \bibinfo {author} {\bibfnamefont {D.}~\bibnamefont {Korshunov}}, \
  and\ \bibinfo {author} {\bibfnamefont {S.}~\bibnamefont {Zachary}},\
  }\href@noop {} {\emph {\bibinfo {title} {An introduction to heavy-tailed and
  subexponential distributions}}},\ Vol.~\bibinfo {volume} {6}\ (\bibinfo
  {publisher} {Springer},\ \bibinfo {year} {2011})\BibitemShut {NoStop}%
\bibitem [{Note1()}]{Note1}%
  \BibitemOpen
  \bibinfo {note} {That could either be a fat-tailed/large-kurtosis
  distribution or special distributions such as Weibull or Gumbel}\BibitemShut
  {NoStop}%
\bibitem [{\citenamefont {Greer}\ \emph {et~al.}(2005)\citenamefont {Greer},
  \citenamefont {Oliver},\ and\ \citenamefont {Nix}}]{greer2005size}%
  \BibitemOpen
  \bibfield  {author} {\bibinfo {author} {\bibfnamefont {J.~R.}\ \bibnamefont
  {Greer}}, \bibinfo {author} {\bibfnamefont {W.~C.}\ \bibnamefont {Oliver}}, \
  and\ \bibinfo {author} {\bibfnamefont {W.~D.}\ \bibnamefont {Nix}},\
  }\href@noop {} {\bibfield  {journal} {\bibinfo  {journal} {Acta Materialia}\
  }\textbf {\bibinfo {volume} {53}},\ \bibinfo {pages} {1821} (\bibinfo {year}
  {2005})}\BibitemShut {NoStop}%
\bibitem [{\citenamefont {Lebensohn}\ and\ \citenamefont
  {Tom{\'e}}(1994)}]{Lebensohn:1994zl}%
  \BibitemOpen
  \bibfield  {author} {\bibinfo {author} {\bibfnamefont {R.}~\bibnamefont
  {Lebensohn}}\ and\ \bibinfo {author} {\bibfnamefont {C.}~\bibnamefont
  {Tom{\'e}}},\ }\href@noop {} {\bibfield  {journal} {\bibinfo  {journal}
  {Materials Science and Engineering: A}\ }\textbf {\bibinfo {volume} {175}},\
  \bibinfo {pages} {71} (\bibinfo {year} {1994})}\BibitemShut {NoStop}%
\bibitem [{\citenamefont {Talamali}\ \emph
  {et~al.}(2011{\natexlab{b}})\citenamefont {Talamali}, \citenamefont
  {Pet{\"a}j{\"a}}, \citenamefont {Vandembroucq},\ and\ \citenamefont
  {Roux}}]{Talamali:2011ij}%
  \BibitemOpen
  \bibfield  {author} {\bibinfo {author} {\bibfnamefont {M.}~\bibnamefont
  {Talamali}}, \bibinfo {author} {\bibfnamefont {V.}~\bibnamefont
  {Pet{\"a}j{\"a}}}, \bibinfo {author} {\bibfnamefont {D.}~\bibnamefont
  {Vandembroucq}}, \ and\ \bibinfo {author} {\bibfnamefont {S.}~\bibnamefont
  {Roux}},\ }\href@noop {} {\bibfield  {journal} {\bibinfo  {journal} {Physical
  Review E}\ }\textbf {\bibinfo {volume} {84}},\ \bibinfo {pages} {016115}
  (\bibinfo {year} {2011}{\natexlab{b}})}\BibitemShut {NoStop}%
\bibitem [{\citenamefont {Zaiser}\ and\ \citenamefont
  {Moretti}(2005)}]{zaiser2005fluctuation}%
  \BibitemOpen
  \bibfield  {author} {\bibinfo {author} {\bibfnamefont {M.}~\bibnamefont
  {Zaiser}}\ and\ \bibinfo {author} {\bibfnamefont {P.}~\bibnamefont
  {Moretti}},\ }\href@noop {} {\bibfield  {journal} {\bibinfo  {journal}
  {Journal of Statistical Mechanics: Theory and Experiment}\ }\textbf {\bibinfo
  {volume} {2005}},\ \bibinfo {pages} {P08004} (\bibinfo {year}
  {2005})}\BibitemShut {NoStop}%
\bibitem [{\citenamefont {Scoville}(2015)}]{Scoville:2015fk}%
  \BibitemOpen
  \bibfield  {author} {\bibinfo {author} {\bibfnamefont {J.}~\bibnamefont
  {Scoville}},\ }\href@noop {} {\bibfield  {journal} {\bibinfo  {journal}
  {arXiv preprint arXiv:1504.01436}\ } (\bibinfo {year} {2015})}\BibitemShut
  {NoStop}%
\bibitem [{\citenamefont {Mueller}\ \emph {et~al.}(2016)\citenamefont
  {Mueller}, \citenamefont {Kusne},\ and\ \citenamefont
  {Ramprasad}}]{Mueller:2016cr}%
  \BibitemOpen
  \bibfield  {author} {\bibinfo {author} {\bibfnamefont {T.}~\bibnamefont
  {Mueller}}, \bibinfo {author} {\bibfnamefont {A.~G.}\ \bibnamefont {Kusne}},
  \ and\ \bibinfo {author} {\bibfnamefont {R.}~\bibnamefont {Ramprasad}},\
  }\href@noop {} {\bibfield  {journal} {\bibinfo  {journal} {Reviews in
  Computational Chemistry}\ }\textbf {\bibinfo {volume} {29}},\ \bibinfo
  {pages} {186} (\bibinfo {year} {2016})}\BibitemShut {NoStop}%
\bibitem [{\citenamefont {Press}\ \emph {et~al.}(1987)\citenamefont {Press},
  \citenamefont {Flannery}, \citenamefont {Teukolsky}, \citenamefont
  {Vetterling},\ and\ \citenamefont {Kramer}}]{Press:1987oq}%
  \BibitemOpen
  \bibfield  {author} {\bibinfo {author} {\bibfnamefont {W.~H.}\ \bibnamefont
  {Press}}, \bibinfo {author} {\bibfnamefont {B.~P.}\ \bibnamefont {Flannery}},
  \bibinfo {author} {\bibfnamefont {S.~A.}\ \bibnamefont {Teukolsky}}, \bibinfo
  {author} {\bibfnamefont {W.~T.}\ \bibnamefont {Vetterling}}, \ and\ \bibinfo
  {author} {\bibfnamefont {P.~B.}\ \bibnamefont {Kramer}},\ }\href@noop {}
  {\enquote {\bibinfo {title} {Numerical recipes: the art of scientific
  computing},}\ } (\bibinfo {year} {1987})\BibitemShut {NoStop}%
\bibitem [{\citenamefont {Greenewald}\ and\ \citenamefont
  {Hero}(2015)}]{Greenewald:2015nx}%
  \BibitemOpen
  \bibfield  {author} {\bibinfo {author} {\bibfnamefont {K.}~\bibnamefont
  {Greenewald}}\ and\ \bibinfo {author} {\bibfnamefont {A.~O.}\ \bibnamefont
  {Hero}},\ }\href@noop {} {\bibfield  {journal} {\bibinfo  {journal} {IEEE
  Transactions on Signal Processing}\ }\textbf {\bibinfo {volume} {63}},\
  \bibinfo {pages} {6368} (\bibinfo {year} {2015})}\BibitemShut {NoStop}%
\bibitem [{\citenamefont {Hinton}\ and\ \citenamefont
  {Salakhutdinov}(2006)}]{dl1}%
  \BibitemOpen
  \bibfield  {author} {\bibinfo {author} {\bibfnamefont {G.~E.}\ \bibnamefont
  {Hinton}}\ and\ \bibinfo {author} {\bibfnamefont {R.~R.}\ \bibnamefont
  {Salakhutdinov}},\ }\href@noop {} {\bibfield  {journal} {\bibinfo  {journal}
  {Science}\ }\textbf {\bibinfo {volume} {313}},\ \bibinfo {pages} {504}
  (\bibinfo {year} {2006})}\BibitemShut {NoStop}%
\bibitem [{\citenamefont {Goodfellow}\ \emph {et~al.}(2014)\citenamefont
  {Goodfellow}, \citenamefont {Pouget-Abadie}, \citenamefont {Mirza},
  \citenamefont {Xu}, \citenamefont {Warde-Farley}, \citenamefont {Ozair},
  \citenamefont {Courville},\ and\ \citenamefont {Bengio}}]{dl2}%
  \BibitemOpen
  \bibfield  {author} {\bibinfo {author} {\bibfnamefont {I.}~\bibnamefont
  {Goodfellow}}, \bibinfo {author} {\bibfnamefont {J.}~\bibnamefont
  {Pouget-Abadie}}, \bibinfo {author} {\bibfnamefont {M.}~\bibnamefont
  {Mirza}}, \bibinfo {author} {\bibfnamefont {B.}~\bibnamefont {Xu}}, \bibinfo
  {author} {\bibfnamefont {D.}~\bibnamefont {Warde-Farley}}, \bibinfo {author}
  {\bibfnamefont {S.}~\bibnamefont {Ozair}}, \bibinfo {author} {\bibfnamefont
  {A.}~\bibnamefont {Courville}}, \ and\ \bibinfo {author} {\bibfnamefont
  {Y.}~\bibnamefont {Bengio}},\ }\href@noop {} {\bibfield  {journal} {\bibinfo
  {journal} {Advances in neural information processing systems}\ ,\ \bibinfo
  {pages} {2672}} (\bibinfo {year} {2014})}\BibitemShut {NoStop}%
\bibitem [{\citenamefont {Dahmen}\ \emph {et~al.}(2009)\citenamefont {Dahmen},
  \citenamefont {Ben-Zion},\ and\ \citenamefont {Uhl}}]{Dahmen:2009kl}%
  \BibitemOpen
  \bibfield  {author} {\bibinfo {author} {\bibfnamefont {K.~A.}\ \bibnamefont
  {Dahmen}}, \bibinfo {author} {\bibfnamefont {Y.}~\bibnamefont {Ben-Zion}}, \
  and\ \bibinfo {author} {\bibfnamefont {J.~T.}\ \bibnamefont {Uhl}},\
  }\href@noop {} {\bibfield  {journal} {\bibinfo  {journal} {Physical review
  letters}\ }\textbf {\bibinfo {volume} {102}},\ \bibinfo {pages} {175501}
  (\bibinfo {year} {2009})}\BibitemShut {NoStop}%
\bibitem [{\citenamefont {Eshelby}(1957)}]{Eshelby:1957fv}%
  \BibitemOpen
  \bibfield  {author} {\bibinfo {author} {\bibfnamefont {J.~D.}\ \bibnamefont
  {Eshelby}},\ }\href@noop {} {\bibfield  {journal} {\bibinfo  {journal}
  {Proceedings of the Royal Society of London. Series A. Mathematical and
  Physical Sciences}\ }\textbf {\bibinfo {volume} {241}},\ \bibinfo {pages}
  {376} (\bibinfo {year} {1957})}\BibitemShut {NoStop}%
\bibitem [{\citenamefont {Budrikis}\ and\ \citenamefont
  {Zapperi}(2013)}]{Budrikis:2013fk}%
  \BibitemOpen
  \bibfield  {author} {\bibinfo {author} {\bibfnamefont {Z.}~\bibnamefont
  {Budrikis}}\ and\ \bibinfo {author} {\bibfnamefont {S.}~\bibnamefont
  {Zapperi}},\ }\href@noop {} {\bibfield  {journal} {\bibinfo  {journal} {arXiv
  preprint arXiv:1307.2135}\ } (\bibinfo {year} {2013})}\BibitemShut {NoStop}%
\bibitem [{\citenamefont {Papanikolaou}(2016)}]{Papanikolaou:2016rt}%
  \BibitemOpen
  \bibfield  {author} {\bibinfo {author} {\bibfnamefont {S.}~\bibnamefont
  {Papanikolaou}},\ }\href@noop {} {\bibfield  {journal} {\bibinfo  {journal}
  {Physical Review E}\ }\textbf {\bibinfo {volume} {93}},\ \bibinfo {pages}
  {032610} (\bibinfo {year} {2016})}\BibitemShut {NoStop}%
\bibitem [{\citenamefont {Kuntz}\ \emph {et~al.}(1998)\citenamefont {Kuntz},
  \citenamefont {Perkovic}, \citenamefont {Dahmen}, \citenamefont {Roberts},\
  and\ \citenamefont {Sethna}}]{Kuntz:1998vl}%
  \BibitemOpen
  \bibfield  {author} {\bibinfo {author} {\bibfnamefont {M.~C.}\ \bibnamefont
  {Kuntz}}, \bibinfo {author} {\bibfnamefont {O.}~\bibnamefont {Perkovic}},
  \bibinfo {author} {\bibfnamefont {K.~A.}\ \bibnamefont {Dahmen}}, \bibinfo
  {author} {\bibfnamefont {B.~W.}\ \bibnamefont {Roberts}}, \ and\ \bibinfo
  {author} {\bibfnamefont {J.~P.}\ \bibnamefont {Sethna}},\ }\href@noop {}
  {\bibfield  {journal} {\bibinfo  {journal} {arXiv preprint cond-mat/9809122}\
  } (\bibinfo {year} {1998})}\BibitemShut {NoStop}%
\bibitem [{\citenamefont {Kuntz}\ \emph {et~al.}(1999)\citenamefont {Kuntz},
  \citenamefont {Perkovic}, \citenamefont {Dahmen}, \citenamefont {Roberts},\
  and\ \citenamefont {Sethna}}]{Kuntz:1999zp}%
  \BibitemOpen
  \bibfield  {author} {\bibinfo {author} {\bibfnamefont {M.~C.}\ \bibnamefont
  {Kuntz}}, \bibinfo {author} {\bibfnamefont {O.}~\bibnamefont {Perkovic}},
  \bibinfo {author} {\bibfnamefont {K.~A.}\ \bibnamefont {Dahmen}}, \bibinfo
  {author} {\bibfnamefont {B.~W.}\ \bibnamefont {Roberts}}, \ and\ \bibinfo
  {author} {\bibfnamefont {J.~P.}\ \bibnamefont {Sethna}},\ }\href@noop {}
  {\bibfield  {journal} {\bibinfo  {journal} {Computing in science \&
  engineering}\ }\textbf {\bibinfo {volume} {1}},\ \bibinfo {pages} {73}
  (\bibinfo {year} {1999})}\BibitemShut {NoStop}%
\bibitem [{\citenamefont {Baeza-Yates}\ and\ \citenamefont
  {Ribeiro-Neto}(2011)}]{f2score}%
  \BibitemOpen
  \bibfield  {author} {\bibinfo {author} {\bibfnamefont {R.}~\bibnamefont
  {Baeza-Yates}}\ and\ \bibinfo {author} {\bibfnamefont {B.}~\bibnamefont
  {Ribeiro-Neto}},\ }\href@noop {} {\emph {\bibinfo {title} {Modern Information
  Retrieval}}},\ Vol.\ \bibinfo {volume} {pp. 327-328.}\ (\bibinfo  {publisher}
  {Addison Wesley},\ \bibinfo {year} {2011})\BibitemShut {NoStop}%
\bibitem [{\citenamefont {Weibull}(1951)}]{Weibull:1951bh}%
  \BibitemOpen
  \bibfield  {author} {\bibinfo {author} {\bibfnamefont {W.}~\bibnamefont
  {Weibull}},\ }\href@noop {} {\bibfield  {journal} {\bibinfo  {journal}
  {Journal of applied mechanics}\ }\textbf {\bibinfo {volume} {18}},\ \bibinfo
  {pages} {293} (\bibinfo {year} {1951})}\BibitemShut {NoStop}%
\bibitem [{\citenamefont {Bouchaud}\ and\ \citenamefont
  {M{\'e}zard}(1997)}]{Bouchaud:1997tg}%
  \BibitemOpen
  \bibfield  {author} {\bibinfo {author} {\bibfnamefont {J.-P.}\ \bibnamefont
  {Bouchaud}}\ and\ \bibinfo {author} {\bibfnamefont {M.}~\bibnamefont
  {M{\'e}zard}},\ }\href@noop {} {\bibfield  {journal} {\bibinfo  {journal}
  {Journal of Physics A: Mathematical and General}\ }\textbf {\bibinfo {volume}
  {30}},\ \bibinfo {pages} {7997} (\bibinfo {year} {1997})}\BibitemShut
  {NoStop}%
\bibitem [{\citenamefont {Le~Doussal}\ and\ \citenamefont
  {Wiese}(2009)}]{Le-Doussal:2009sp}%
  \BibitemOpen
  \bibfield  {author} {\bibinfo {author} {\bibfnamefont {P.}~\bibnamefont
  {Le~Doussal}}\ and\ \bibinfo {author} {\bibfnamefont {K.~J.}\ \bibnamefont
  {Wiese}},\ }\href@noop {} {\bibfield  {journal} {\bibinfo  {journal}
  {Physical Review E}\ }\textbf {\bibinfo {volume} {79}},\ \bibinfo {pages}
  {051105} (\bibinfo {year} {2009})}\BibitemShut {NoStop}%
\bibitem [{\citenamefont {Papanikolaou}\ \emph {et~al.}(2015)\citenamefont
  {Papanikolaou}, \citenamefont {Song},\ and\ \citenamefont {Van~der
  Giessen}}]{Papanikolaou:2015wt}%
  \BibitemOpen
  \bibfield  {author} {\bibinfo {author} {\bibfnamefont {S.}~\bibnamefont
  {Papanikolaou}}, \bibinfo {author} {\bibfnamefont {H.}~\bibnamefont {Song}},
  \ and\ \bibinfo {author} {\bibfnamefont {E.}~\bibnamefont {Van~der
  Giessen}},\ }\href@noop {} {\bibfield  {journal} {\bibinfo  {journal} {J.
  Mech. Phys. Solids.}\ }\textbf {\bibinfo {volume} {102}},\ \bibinfo {pages}
  {17} (\bibinfo {year} {2015})}\BibitemShut {NoStop}%
\bibitem [{\citenamefont {Parthasarathy}\ \emph {et~al.}(2007)\citenamefont
  {Parthasarathy}, \citenamefont {Rao}, \citenamefont {Dimiduk}, \citenamefont
  {Uchic},\ and\ \citenamefont {Trinkle}}]{Parthasarathy:2007fk}%
  \BibitemOpen
  \bibfield  {author} {\bibinfo {author} {\bibfnamefont {T.~A.}\ \bibnamefont
  {Parthasarathy}}, \bibinfo {author} {\bibfnamefont {S.~I.}\ \bibnamefont
  {Rao}}, \bibinfo {author} {\bibfnamefont {D.~M.}\ \bibnamefont {Dimiduk}},
  \bibinfo {author} {\bibfnamefont {M.~D.}\ \bibnamefont {Uchic}}, \ and\
  \bibinfo {author} {\bibfnamefont {D.~R.}\ \bibnamefont {Trinkle}},\
  }\href@noop {} {\bibfield  {journal} {\bibinfo  {journal} {Scripta
  Materialia}\ }\textbf {\bibinfo {volume} {56}},\ \bibinfo {pages} {313}
  (\bibinfo {year} {2007})}\BibitemShut {NoStop}%
\bibitem [{\citenamefont {Tadmor}\ \emph {et~al.}(2011)\citenamefont {Tadmor},
  \citenamefont {Elliott}, \citenamefont {Sethna}, \citenamefont {Miller},\
  and\ \citenamefont {Becker}}]{Tadmor:2011ty}%
  \BibitemOpen
  \bibfield  {author} {\bibinfo {author} {\bibfnamefont {E.}~\bibnamefont
  {Tadmor}}, \bibinfo {author} {\bibfnamefont {R.}~\bibnamefont {Elliott}},
  \bibinfo {author} {\bibfnamefont {J.~P.}\ \bibnamefont {Sethna}}, \bibinfo
  {author} {\bibfnamefont {R.}~\bibnamefont {Miller}}, \ and\ \bibinfo {author}
  {\bibfnamefont {C.~A.}\ \bibnamefont {Becker}},\ }\href@noop {} {\bibfield
  {journal} {\bibinfo  {journal} {JOM Journal of the Minerals, Metals and
  Materials Society}\ }\textbf {\bibinfo {volume} {63}},\ \bibinfo {pages} {17}
  (\bibinfo {year} {2011})}\BibitemShut {NoStop}%
\bibitem [{\citenamefont {Frenkel}\ and\ \citenamefont
  {Smit}(2002)}]{Frenkel:2002qv}%
  \BibitemOpen
  \bibfield  {author} {\bibinfo {author} {\bibfnamefont {D.}~\bibnamefont
  {Frenkel}}\ and\ \bibinfo {author} {\bibfnamefont {B.}~\bibnamefont {Smit}},\
  }\href@noop {} {\emph {\bibinfo {title} {Understanding molecular simulation:
  from algorithms to applications}}},\ Vol.~\bibinfo {volume} {1}\ (\bibinfo
  {publisher} {Elsevier (formerly published by Academic Press)},\ \bibinfo
  {year} {2002})\BibitemShut {NoStop}%
\bibitem [{\citenamefont {Lloyd}(1982)}]{Lloyd:1982hb}%
  \BibitemOpen
  \bibfield  {author} {\bibinfo {author} {\bibfnamefont {S.}~\bibnamefont
  {Lloyd}},\ }\href@noop {} {\bibfield  {journal} {\bibinfo  {journal} {IEEE
  transactions on information theory}\ }\textbf {\bibinfo {volume} {28}},\
  \bibinfo {pages} {129} (\bibinfo {year} {1982})}\BibitemShut {NoStop}%
\bibitem [{\citenamefont {Arthur}\ and\ \citenamefont
  {Vassilvitskii}(2006)}]{Arthur:2006kl}%
  \BibitemOpen
  \bibfield  {author} {\bibinfo {author} {\bibfnamefont {D.}~\bibnamefont
  {Arthur}}\ and\ \bibinfo {author} {\bibfnamefont {S.}~\bibnamefont
  {Vassilvitskii}}\ }(\bibinfo  {publisher} {ACM},\ \bibinfo {year} {2006})\
  pp.\ \bibinfo {pages} {144--153}\BibitemShut {NoStop}%
\end{thebibliography}
%

\vspace{0.5cm}
{\bf Acknowledgements}\\
{  D. M. Dimiduk, }H. Song, H. Yavas, E.~Van~der~Giessen are gratefully thanked for inspiring discussions. This work is supported by the US Department of Energy, Award Number  DE-SC0014109, and by the Department of Commerce - NIST, Award Number 1007294R.

\vspace{0.5cm}
{\bf Competing interests}\\
The author declares no competing financial interests.

\vspace{0.5cm}
{\bf Author information}\\
{\bf Affiliations}\\
{Department of Mechanical\&Aerospace Engineering, The West Virginia University}\\
{Department of Physics, The West Virginia University}\\
{Department of Mechanical Engineering, The Johns Hopkins University, Baltimore, MD 21218}

\end{document}